\documentstyle[aps,prb]{revtex}
\newcommand{\supscrpt}[2]{{#1}^{{#2}}}
\newcommand{\subscrpt}[2]{{#1}_{{#2}}}
\newcommand{\mixten}[3]{{#1}^{#2}_{\phantom{{#2}} #3}}
\newcommand{\supsub}[3]{{#1}^{#2}_{#3}}

\newcommand{\bfc}{{\bf c}}
\newcommand{\bfcsup}[1]{\supscrpt{\bfc}{#1}}
\newcommand{\bfci}{\bfcsup{i}}

\newcommand{\bfx}{{\bf x}}

\newcommand{\bfxt}{(\bfx, t)}
\newcommand{\bfy}{{\bf y}}

\newcommand{\bge}{\begin{equation}}
\newcommand{\ee}{\end{equation}}
\newcommand{\bgc}{\begin{center}}
\newcommand{\ec}{\end{center}}
\newcommand{\bgea}{\begin{eqnarray}}
\newcommand{\eea}{\end{eqnarray}}
\newcommand{\bgeas}{\begin{eqnarray*}}
\newcommand{\eeas}{\end{eqnarray*}}
\newtheorem{thm}{Theorem}

\newcommand{\advop}[2]{\mixten{{\cal A}}{#1}{#2}}

\newcommand{\jop}[2]{\mixten{J}{#1}{#2}}
\newcommand{\jij}{\jop{i}{j}}

\newcommand{\tjop}[2]{\mixten{\tilde{J}}{#1}{#2}}
\newcommand{\tjij}{\tjop{i}{j}}

\newcommand{\cjop}[2]{\mixten{{\cal J}}{#1}{#2}}
\newcommand{\kop}[2]{\mixten{K}{#1}{#2}}
\newcommand{\ckop}[2]{\mixten{{\cal K}}{#1}{#2}}

\newcommand{\vop}[2]{\mixten{V}{#1}{#2}}
\newcommand{\ctop}[2]{\mixten{{\cal T}}{#1}{#2}}
\newcommand{\cvop}[2]{\mixten{{\cal V}}{#1}{#2}}

\newcommand{\csup}[1]{\supscrpt{\omega}{#1}}
\newcommand{\ccsup}[1]{\supscrpt{\Omega}{#1}}

\newcommand{\fsup}[1]{\supscrpt{f}{#1}}
\newcommand{\gsup}[1]{\supscrpt{g}{#1}}
\newcommand{\nsup}[1]{\supscrpt{n}{#1}}
\newcommand{\nixt}{\nsup{i}\bfxt}
\newcommand{\nnsup}[1]{\supscrpt{N}{#1}}
\newcommand{\nnsupo}[2]{\supsub{N}{#1}{#2}}

\newcommand{\ppsup}[1]{\supscrpt{P}{#1}}

\newcommand{\ggamsup}[1]{\supscrpt{\Gamma}{#1}}
 \newcommand{\alpsupp}[1]{\supscrpt{\alpha}{#1}}
 \newcommand{\betsupp}[1]{\supscrpt{\beta}{#1}}

\newcommand{\phisup}[1]{\supscrpt{\Phi}{#1}}

\newcommand{\pdv}{\partial}
\newcommand{\dt}{\Delta t}
\newcommand{\tpdt}{t+\dt}

\def\glossitem#1!#2!#3{$#1$ \> \if !#2! \else \pageref{#2} \fi\>
    \parbox[t]{3.5truein}{#3} \\}

\def\registeredsymbol{
  {\ooalign{\hfil\raise.05ex\hbox{\hskip.02ex$\rm
   \scriptscriptstyle R$}\hfil\crcr
   \hbox{$\scriptstyle\mathchar"20D$}}}}
\def\registered{
  \ifmmode^\registeredsymbol\else$^\registeredsymbol$\fi}

\def\vertex#1#2#3#4{\begin{picture}(8,8)(- 4,- 4)
\put(0,0){\circle*{3}}
\ifnum#1=1 \put(0,0){\line(1,0){4}}\fi
\ifnum#2=1 \put(0,0){\line(0,1){4}}\fi
\ifnum#3=1 \put(0,0){\line(-1,0){4}}\fi
\ifnum#4=1 \put(0,0){\line(0,-1){4}}\fi
\end{picture}}

\begin{document}

\title{\begin{flushleft}
                      {\footnotesize BU-CCS-941001, MIT-CTP-2280}\\
                      {\footnotesize comp-gas/9411001}\\
                      {\footnotesize To appear, {\it Proceedings of the
Workshop on Pattern Formation and Lattice Gases}, Fields Institute
(1995)}\\[0.3in]
                      \end{flushleft}
Renormalization of Lattice Gas Transport Coefficients\footnote{
\small \baselineskip=11pt This work was supported in part by Thinking
Machines Corporation, in part by the divisions of Applied Mathematics of
the U.S. Department of Energy (DOE) under contracts DE-FG02-88ER25065
and DE-FG02-88ER25066, and in part by the U.S. Department of Energy
(DOE) under cooperative agreement DE-FC02-94ER40818.}
}
\author{
Bruce M. Boghosian\\
{\small \sl Center for Computational Science,}\\
{\small \sl Boston University,}\\
{\small \sl 3 Cummington Street, Boston, Massachusetts 02215, U.S.A.} \\
{\small \tt bruceb@conx.bu.edu} \\
[0.3cm]
Washington Taylor\\
{\small \sl Center for Theoretical Physics,}\\
{\small \sl Laboratory for Nuclear
Science and Department of Physics,}\\
{\small \sl Massachusetts Institute of
Technology; Cambridge, Massachusetts 02139, U.S.A.} \\
{\small \tt wati@mit.edu} \\
}
\date{(\today)}
\maketitle

\begin{abstract}
A method is described for calculating corrections to the
Boltzmann/Chapman-Enskog analysis of lattice gases due to the buildup of
correlations.  It is shown that renormalized transport coefficients can
be calculated perturbatively by summing terms in an infinite series.  A
diagrammatic notation for the terms in this series is given, in analogy
with the Feynman diagrams of quantum field theory.  This theory is
applied to an example lattice gas and shown to correctly predict
experimental deviation from the Boltzmann prediction.
\end{abstract}

\section{Introduction}
\setcounter{equation}{0}

The central problem in the theoretical analysis of a lattice gas is the
determination of the macroscopic hydrodynamic equations obeyed by the
conserved quantities of the system, with transport coefficients
expressed as functions of those conserved quantities.  Most analyses of
this problem make use of the Boltzmann molecular chaos assumption.  This
assumption, which neglects correlations between colliding particles,
makes possible the derivation of a Boltzmann equation for the
single-particle distribution function.  For lattice gases satisfying
semi-detailed balance, this Boltzmann equation has Fermi-Dirac
equilibria.  The Chapman-Enskog procedure of linearizing about these
equilibria then yields the hydrodynamic equations satisfied by the
system.  The transport coefficients that appear in these equations are
related to the eigenvalues of the linearized Boltzmann collision
operator at equilibrium, and these are calculable as functions of the
conserved quantities.

Unfortunately, the first step in this analysis -- the Boltzmann
molecular chaos assumption -- is often invalid, and its use can lead to
serious errors in the transport coefficients, as has been shown
empirically \cite{bmbcdltdfp}.  To see why, recall that the transport
coefficients measure the rate at which the propagation/collision process
smoothes gradients.  These same collisions, however, generate
correlations among the bits of the lattice gas, and these correlations
can propagate about and subsequently correct the Boltzmann collision
operator.  Thus, in an exact theory, the {\it renormalized} linearized
collision operator is the sum of a {\it bare} Boltzmann collision
operator, and a series of corrections due to the presence of
correlations.

For example, consider an event in which two particles emerge from a
collision (and thereby acquire a correlation), move about in a
background of uncorrelated particles, and later recollide.  This process
can be thought of as a one-loop correction to the Boltzmann
approximation.  Further refinements can be obtained by including more
intricate diagrams -- with multiple loops, nested loops, etc. -- to
account for the interaction of the correlated quantities with the
background.  Standard field-theoretic techniques can then be used to
approximate these diagrammatic sums.

In this work, we outline a systematic approach to the exact analysis of
a general lattice gas.  We show how to construct a diagrammatic series
for the transport coefficients of any lattice gas satisfying
semi-detailed balance\footnote{at lowest order}, and we provide
closed-form expressions for the vertex coefficients.  These results
constitute the central part of this paper and are contained in Section
3.  The methods that we use here are analogous to the cluster expansion
techniques that have been used for many years in the analysis of
continuum fluids~\cite{kringfl}.  The discrete nature of lattice gases,
however, necessitates a construction which differs substantially from
the continuum analysis.

The formalism described here is {\it exact}, in that it yields the
transport coefficients with no approximation.  Of course, it is
usually a very difficult combinatorial task to sum the diagrammatic
series involved, and, more often than not, one must content oneself
with an approximation to this series.  In doing so, however, it is
possible to make contact with other methods of computing this
correction.  In particular, it is possible to understand kinetic ring
theory, and BBGKY truncations in terms of truncations of the exact
diagrammatic series, and to characterize the classes of diagrams that
they retain, and those that they ignore.

Our analysis differs from previous lattice gas analyses in a few
important ways.  We formulate the theory in terms of connected
correlation functions.  A similar expansion in terms of products of
fluctuations was previously used in~\cite{Ern,ernstd} and
in~\cite{vanvel,kringlob}, where the ring kinetic theory was derived for
lattice gases, and for lattice Lorentz gases, respectively.  The use of
the connected correlation functions simplifies the form of the complete
diagrammatic expansion for the kinetic theory significantly.  Previous
work on the kinetic ring theory of lattice gases has used the Green-Kubo
formalism to obtain the series for the transport coefficients.  In this
paper, we use the Chapman-Enskog theory instead.  In this way, we get
the Boltzmann approximation at zeroth order.  All higher-order terms in
our series are thus corrections to the Boltzmann approximation.  In the
Green-Kubo theory, by contrast, it is necessary to sum an infinite
number of terms just to get the Boltzmann approximation.

We describe in Section 4 a variety of simplifications and approximations
which can be made to simplify the numerical computation of the
renormalized transport coefficients.  The simplest of these
approximations is the ring approximation, in which correlations between
more than 2 particles are neglected, and correlations are not allowed to
interact.  Thus, the results of~\cite{Ern} are described in our theory
by the restriction of the diagrammatic sum to the simple set of ring
diagrams.

Finally, to make all this more concrete, in Section 5 a worked example
is given, and the results are compared with experiment.

\section{Definitions and Notation}
\setcounter{equation}{0}

A lattice gas is generally described by a state space and a
time-development rule.  The state space is defined by associating $n$
bits with each point on a lattice $L$.  (Bits are variables taking
values in $\{0,1\}$.)  We define the set of bits at a general lattice
site to be $B$, so that $|B|=n$.  We denote the total number of bits on
the lattice by $N = n|L|$.  For each value of the discrete time
parameter $t$, we write the values of the bits as $\nixt$, where
$\bfx\in L$ and $i\in B$.  These bits can be thought of as a set of
occupation numbers for individual {\sl particle} states.  Each of the
bits $n^i$ is associated with a lattice vector $\bfc^i$, which
represents the velocity vector of the associated particle.

We define the set $S$ of possible states of the bits at a general
lattice point at a fixed value of $t$ to be
\[
   S=\{s: s\subseteq B\},
\]
where a state $s$ is associated with the set of bits taking the value
$1$.
Note that $|S|=2^n$.  \label{pg:al}

We shall often want to refer to the $N=n|L|$ bits of the lattice in a
uniform fashion, so we introduce an enumeration of these $N$ bits, given
by a 1-1 correspondence between the sets ${\cal B}= \{1, 2, \ldots, N\}$ and
$B \times L = \{(i,\bfx): i\in B,\bfx\in L\}$.  In this notation, a single
bit of the lattice gas is written as $\nsup{a}$, $a\in{\cal B} $.
To relate this notation to the more explicit $(i,\bfx)$ notation, we
express the above 1-1 correspondence by writing $a$ and $(i,\bfx)$ as
functions of one another, so that
\[
   \nsup{a}(t) = \nsup{i(a)} (\bfx (a), t)
\]
and
\[
   \nixt = \nsup{a(i,\bfx)}(t).
\]
We shall use both notations interchangeably throughout this paper.

\section{The Renormalized Chapman-Enskog Analysis}
\setcounter{equation}{0}
\subsection{Ensemble averages}

We will denote the mean of an arbitrary product of the $\nsup{a}$'s by
\[
 \nnsup{\alpha} = \left\langle\prod_{a\in\alpha}\nsup{a}\right\rangle, \;\;
 \alpha \subseteq \{1, \ldots, N\}.
\]
There are $2^N-1$ independent $\nnsup{\alpha}$'s, since
$\nnsup{\emptyset} = \langle 1 \rangle = 1$.  We will sometimes use a
roman index to denote an index set with a single element, as in
$\nnsup{a}=\nnsup{\{a\}}$.

We introduce an advection operator, $\advop{\alpha}{\beta}$, which is a
permutation matrix on the $2^N-1$ dimensional space of means.
$\advop{\alpha}{\beta}$ takes a set of particles $\beta = \{b_1, \ldots,
b_q\}$ to the advected set of particles $\alpha = \{a_1, \ldots, a_q\}$,
where $a_j=a(i(b_j), \bfx(b_j) + \bfc^{i(b_j)})$.  Note that this
generalizes the usual concept of an advection operator, that propagates
single particles from site to site, to one that can propagate arbitrary
multipoint quantities (e.g., connected correlation functions) from site
to site.

\subsection{Boltzmann Analysis}

The evolution of a lattice gas in one time step can be divided into two
parts.  The first part is a {\it collision} phase in which the $n$ bits
at each point alter their values in some specified way.  The second part
is a {\it propagation} phase in which the value of each bit moves to the
corresponding bit of the lattice point to which it is carried by its
associated velocity vector.  The complete time-development equation is
then written as
\bge
        \nsup{i} (\bfx + \bfcsup{i}, \tpdt)=
	\nixt + \csup{i}\left(\nsup{\ast}\bfxt\right).
   \label{eq:microdyn}
\ee
The {\it collision operator} $\csup{i}$ is generally a nonlinear
function of its arguments.  Note that the form $f(z^\ast)$ is used to
indicate that a function $f$ depends on a quantity $z$ for all possible
values of the index $*$.

The Boltzmann molecular chaos assumption can be written
\[
\langle\csup{i}\left(\nsup{\ast}\bfxt\right)\rangle \approx
\csup{i}\left(\left\langle \nsup{\ast}\bfxt\right\rangle\right).
\]
This is inexact when $\csup{i}$ is nonlinear.

Taking the ensemble average of Eq.~(\ref{eq:microdyn}), and making the
molecular chaos assumption leads to the Boltzmann equation,
\bge
  \nnsup{b}(\tpdt) =
    \advop{b}{c}
      \left(
        \nnsup{c}\left(t\right)+
        \ccsup{i(c)}\left(
                  \nnsup{a(*,\bfx (c))}\left(t\right)
                \right)
      \right),
\label{eq:microdynm}
\ee
where repeated indices ($c$ in this case) are summed.  This is a closed
system of equations for the $\nnsup{a}$'s.

The usual Chapman-Enskog analysis proceeds by expanding this equation of
motion using Navier-Stokes ordering: A formal expansion parameter
$\epsilon$ is introduced, such that $|\bfci| \sim \epsilon$.  The time
step is ordered as $\dt \sim \epsilon^2$, as is appropriate for viscous
or diffusive processes.  Since the lattice gas is assumed to satisfy
semi-detailed balance, there is a Fermi-Dirac equilibrium.  It is
assumed that deviations from this equilibrium are of order $\epsilon$.
We will denote the equilibrium values of the single-particle means by
$\nnsupo{a}{0}$.

Expanding the collision operator in (\ref{eq:microdynm}) to first
order in $\epsilon$ gives
\bge
\nnsup{b}(\tpdt)=
   \advop{b}{c} \left(N^c (t)+
	\epsilon                      \jop{i(c)}{j}
   \nnsup{a(j,\bfx (c))}_1 (t)\right).
\label{eq:eqfornone}
\ee
where $\jop{i}{j} =
\left.\partial\ccsup{i}(\nnsup{\ast})/\partial\nnsup{j}\right|_0$ is the
Jacobian matrix of the collision operator evaluated at the equilibrium.

The rest of the Chapman-Enskog analysis consists of associating the
hydrodynamic and kinetic modes of the system with eigenvectors of $J$,
and writing the hydrodynamic equations in such a way that the transport
coefficients are functions of the eigenvalues of $J$.  For a more
in-depth review of the details of this procedure, see \cite{bw1}.

\subsection{Exact Analysis}

We will now drop the Boltzmann assumption and show that the exact
equation of motion leads in the scaling limit to an equation identical
to (\ref{eq:eqfornone}), but with a {\it renormalized} $J$ matrix.  We
will expand about the same equilibrium as in the usual analysis.  For
any lattice gas satisfying semi-detailed balance, this equilibrium
exists; in the vicinity of this equilibrium, correlations of at most
order $\epsilon$ are generated in a single time step.

We begin with the exact time-development equation (\ref{eq:microdyn}).
By taking the ensemble average of the product of this equation over all
$a$ in an arbitrary set $\alpha \subseteq {\cal B}$, we can write the
exact equation for an arbitrary multipoint mean at time $\tpdt$ in terms
of multipoint means at time $t$.  We have
\bge
\nnsup{\alpha}(\tpdt) =
\left\langle \prod_{a \in \alpha} \nsup{a} (\tpdt) \right\rangle =
\sum_\beta \advop{\alpha}{\beta}
\left\langle \prod_{b \in \beta}
\left[\nsup{b} (t) + \csup{i(b)}(\nsup{\ast}(\bfx(b), t))\right]
\right\rangle.
\label{eq:exactm}
\ee
To express the right-hand side in terms of multipoint means, it will be
convenient to rewrite this equation in a more compact notation.  For a
set $\beta \subseteq {\cal B}$, let us define $L_\beta$ to be the subset
of points in $L$ which contain at least one particle in the set $\beta$;
that is,
\[
   L_\beta = \{ \bfy \in L : \bfx(b) = \bfy\;\; {\rm for\; some}\;\; b \in
   \beta\}.
\]
Similarly, we define $\beta_\bfx$ to be the set of $i$'s corresponding
to the particles in $\beta$ at the point $\bfx$; that is,
\[
   \beta_\bfx = \{ i \in B: a(i, \bfx) \in\beta\}.
\]
We can now factorize \label{pg:ad} the product appearing on the
right-hand side of (\ref{eq:exactm}) into contributions from each of the
points in $L_\beta$, by writing
\[
   \prod_{b \in \beta}
   \left[\nsup{b} (t) + \csup{i(b)}(\nsup{\ast}(\bfx(b), t))\right] =
   \prod_{\bfx \in L_\beta}
   \prod_{i \in \beta_\bfx}
   \left[\nixt + \csup{i}(\nsup{\ast}\bfxt)\right].
\]
The innermost product on the right now depends only on quantities at a
single site, $\bfx$.

The functions $\csup{i}(\nsup{\ast})$ can be expressed as polynomials in
the $\nsup{i}$'s of the form
\[
   \csup{i}(\nsup{\ast}) = \sum_{\nu \subseteq B} \mixten{k}{i}{\nu}
                            \prod_{j \in \nu} \nsup{j},
\]
where the $\mixten{k}{i}{\nu}$ are coefficients which may depend only on
stochastic variables at each lattice site, and which are constant for
deterministic lattice gases.  \label{pg:bc} Thus, we can write
\[
   \prod_{i \in \mu} [\nsup{i} + \csup{i}(\nsup{\ast})] =
   \sum_{\nu\subseteq B}\mixten{v}{\mu}{\nu} \prod_{j \in \nu}\nsup{j},
\]
where the quantities $\mixten{v}{\mu}{\nu}$ may contain stochastic
variables at each site.  Taking the ensemble average over any such
variables, we get the {\it mean vertex coefficients}
$\mixten{V}{\mu}{\nu}$
\bge
\mixten{V}{\mu}{\nu} = \langle\mixten{v}{\mu}{\nu} \rangle.
\label{eq:vertexmean1}
\ee

The lattice gas state transition probabilities $A (s \rightarrow s')$
may be interpreted as elements of a collision matrix on the space of
probabilities, $\ppsup{s}$, in the sense that the post-collision
probability of a state $s'$ is given by
\bge
 \sum_s A (s \rightarrow s') \ppsup{s}.
 \label{eq:cmosops}
\ee
Similarly, the matrix $\vop{\mu}{\nu}$ can be interpreted as a collision
matrix on the space of means.  Since the probabilities and the
multipoint means are straightforwardly related, it is possible to relate
the matrix $V$ to the transition probabilities.  We get
\bge
 \mixten{V}{\mu}{\nu} = \sum_{s' \supseteq \mu}
 \sum_{s \subseteq \nu}(-1)^{| \nu | - | s |} A (s \rightarrow s'),
 \label{eq:vertexmean2}
\ee
where we have identified the state $s$ with the set of bits which are 1
in that state ($s \subseteq B$).  Clearly, the $2^{2n}$ matrix elements
$\vop{\mu}{\nu}$ depend only on the sets $\mu$ and $\nu$, and on the
form of the collision operator.  In particular, they do not depend on
$\bfx$, or on the values of the $\nsup{a}$'s.  Note that
$\vop{\emptyset}{\nu} = \mixten{\delta}{\emptyset}{\nu}$, regardless of
the specific lattice gas or collision rule.

Eq.~(\ref{eq:exactm}) can now be rewritten in the form
\bge
\nnsup{\alpha}(\tpdt) =
   \advop{\alpha}{\beta} \kop{\beta}{\gamma} \nnsup{\gamma} (t),
   \label{eq:mdabs}
\ee
where $K$ is an operator expressing the complete collision part of the
time development, given by
\bge
   \kop{\beta}{\gamma}
   = \prod_{\bfx \in L_\beta} \vop{\beta_\bfx}{\gamma_\bfx}.
   \label{eq:dmprod}
\ee

We shall now rewrite the dynamical equation (\ref{eq:mdabs}) in the
language of connected correlation functions (CCF's).  For each
$\alpha\subseteq\{1, \ldots, N\}$, there is a CCF, written
$\Gamma^\alpha$.  It is simplest to relate the CCF's to the means by the
equation
\bge
N^\alpha = f^\alpha(\Gamma^\ast) =
   \sum_{\pi(\alpha)} \Gamma^{\zeta_1} \ldots \Gamma^{\zeta_q},
\label{eq:meanccf}
\ee
where $\pi(\alpha)$ is the set of all partitions of $\alpha$ into
disjoint subsets, $\zeta_1, \ldots, \zeta_q$.  Thus, for example, we
have $N^a=\Gamma^a$ and
\[
\nnsup{abc} = \Gamma^{abc} + \Gamma^{a}\Gamma^{bc} + \Gamma^{b}\Gamma^{ac}+
          \Gamma^{c}\Gamma^{ab} + \Gamma^a \Gamma^b \Gamma^c.
\]
For notational convenience, an index set with a circumflex, e.g.
$\hat{\alpha}$, is assumed to have at least two elements.  We refer to
the mean (CCF) of a set with $q$ elements as a $q$-mean ($q$-CCF).

Equation (\ref{eq:meanccf}) can be inverted by induction on $|\alpha|$
to get an expression of the form $\Gamma^\alpha = g^\alpha(N^\ast)$,
where $g = f^{-1}$.  We can now use $f$ and $g$ to convert
(\ref{eq:mdabs}) into a dynamical equation for the CCF's, giving
\bge
   \ggamsup{\alpha}(\tpdt) =
   \advop{\alpha}{\beta} \phisup{\beta}(\ggamsup{\ast}),
   \label{eq:gamtd}
\ee
where
\[
   \phisup{\beta}(\ggamsup{\ast}) \equiv
   \gsup{\beta}(\kop{\ast}{\gamma} \fsup{\gamma}(\ggamsup{\ast}));
\]
we have used the fact that ${\cal A}$ is a permutation matrix, so
that $g$ and ${\cal A}$ commute.

For an ensemble of independent variables, all CCF's except the 1-CCF's
vanish. The equilibrium distribution is such an ensemble, so we can take
any term $\Gamma^{\widehat{\alpha}}$ to be of order $\epsilon$.  In the
asymptotic analysis, only first order terms in $\epsilon$ need be kept
in the expression $\Phi^{\widehat{\beta}}(\Gamma^{\ast})$.  Thus, the dynamical
equation for the CCF's can be linearized to
\bge
   \ggamsup{\widehat{\alpha}}(\tpdt) =
   \advop{\widehat{\alpha}}{\widehat{\beta}} \left(
   \epsilon \ckop{\widehat{\beta}}{a}
   \nnsup{a}_1 (t) +
   \ckop{\widehat{\beta}}{\widehat{\gamma}}
   \ggamsup{\widehat{\gamma}}(t)\right),
   \label{eq:cdlin}
\ee
where \label{pg:bf}
\bge
   \ckop{\beta}{\gamma} = \subscrpt{\left.
   \frac{\pdv \phisup{\beta}}{\pdv \ggamsup{\gamma}}
   \right|}{0} =  \subscrpt{\left.
   \frac{\pdv \gsup{\beta}}{\pdv \nnsup{\sigma}}
   \right|}{0} \kop{\sigma}{\tau} \subscrpt{\left.
   \frac{\pdv \supscrpt{f}{\tau}}{\pdv \ggamsup{\gamma}}
   \right|}{0}.
   \label{eq:dexp}
\ee
Similarly,  when we  include the effects of correlations to order
$\epsilon$ in (\ref{eq:gamtd}) for  $\alpha= \{a\}$,  the dynamical
equation for $N^a = \Gamma^a$ becomes
\bge
\nnsup{a}(\tpdt)  =
   \advop{a}{ b}
     \left\{
        N^b (t)+
	\epsilon\left[ \ckop{b}{c} -\mixten{\delta}{b}{c}\right]
        \nnsup{c}_1 (t) +
        \ckop{b}{\widehat{\gamma}} \ggamsup{\widehat{\gamma}}(t)
     \right\}.
  \label{eq:exact1}
\ee
Note that if we set
$\ggamsup{\widehat{\alpha}} = 0$ in this equation, we get back the
linearized Boltmann equation (\ref{eq:eqfornone}), since
\[
\ckop{b}{c} -\mixten{\delta}{b}{c} = \mixten{\delta}{\bfx(b)}{\bfx(c)}
\mixten{J}{i (b)}{i (c)}.
\]
Inserting (\ref{eq:cdlin}), we can
now write the exact equation of motion
for the quantities $\nnsup{a}$ in the form of an infinite series,
\bge
\nnsup{a}(\tpdt)  =
   \advop{a}{ b} \left(N^b (t)+
	\epsilon \cjop{b}{c}
	\nnsup{c}_1 (t)\right),
  \label{eq:rbolt}
\ee
where
\bge
   \cjop{b}{c} =
\mixten{\delta}{\bfx(b)}{\bfx(c)}
\mixten{J}{i (b)}{i (c)} +
  \ckop{b}{\widehat{\alpha}}
   (\advop{\widehat{\alpha}}{\widehat{\beta}} \ckop{\widehat{\beta}}{c} +
   \advop{\widehat{\alpha}}{\widehat{\beta}}
   \ckop{\widehat{\beta}}{\widehat{\gamma}}
   (\advop{\widehat{\gamma}}{\widehat{\delta}}\ckop{\widehat{\delta}}{c} +
   \ldots)).
   \label{eq:rj}
\ee
\label{pg:bb} We now have an expression for the mean occupation number
of a certain bit of the system at position $\bfx$ and time $t + \Delta
t$, written as an infinite sum of terms, each of which is a function of
the quantities $N_0$ and $N_1$ at nearby lattice sites $\bfx'$ and at
previous time steps $t'$.  As we consider terms in this series with more
and more factors of ${\cal A}{\cal K}$, the positions and times at which
these quantities are evaluated will differ from $\bfx$ and $t$ by
greater amounts.  However, for any given term in the series, the means
$N_0^{a (i,\bfx')}(t')$ and $N_1^{a (i,\bfx')}(t')$ can be replaced by
$N_0^{a (i,\bfx)}(t)$ and $N_1^{a (i,\bfx)}(t)$, and the expression
(\ref{eq:rj}) will only change by a quantity of order $\epsilon$, since
spatial derivatives are ordered as $\epsilon$, and temporal derivatives
are ordered as $\epsilon^2$.  Such a modification for a finite number of
terms does not change the behavior of the system in the hydrodynamic
limit.  In fact, it follows that whenever the sum of terms in
(\ref{eq:rj}) converges on a scale which goes to zero in the
hydrodynamic limit, we can drop all the spatial and temporal variations
in the single-particle means.  Thus, (\ref{eq:rbolt}) can be rewritten
in precisely the form of Eq.~(\ref{eq:eqfornone}), where $J$ is
replaced by the renormalized matrix \label{pg:ba}
\bge
   \tjop{i}{j} (\bfx) =
   \sum_{\bfy\in L}
   \cjop{a(i, \bfx)}{a(j, \bfy)},
   \label{eq:rjt}
\ee
with all ${\cal K}$'s in ${\cal J}$ evaluated at the point $\bfx$ and
the time $t$.

It is important to note that the above argument breaks down when the
sum (\ref{eq:rj}) is divergent.  In this case, the effects of large
scale variations in the $\nnsup{a(i,\bfx)}$'s must be considered.  In
general, for lattice gases where (\ref{eq:rj}) is divergent, one must
be quite careful about the analysis.  For certain lattice gases,
however, particularly systems in which the conserved quantities are
ordered, we are interested in expanding around an equilibrium which is
spatially invariant (for example, the FHP-I lattice gas).  In this
case,  the zero-order means can be replaced by their universal values;
however, one must still treat the spatial variation of the
first-order means carefully.

Now that we have rewritten the exact dynamical equation in a form
commensurate with the original form of the lattice Boltzmann equation,
the renormalized transport coefficients for the theory can be
related to the eigenvalues of the matrix $\tjij$ in the same way that
the original (Boltzmann) transport coefficients were related to the
eigenvalues of the matrix $\jij$.  Thus, if we can compute the matrix
$\tjij$ exactly, we can also compute the exact renormalized transport
coefficients.

We will now show that expression~(\ref{eq:rj}) for ${\cal J}$ can be
written in a diagrammatic notation, allowing us to perform a
perturbative calculation of ${\cal J}$ by summing over ``Feynman
diagram''-like objects, where the contribution from each diagram is just
the product of factors associated with its vertices.  The key
observation which allows this reduction is the following:
\begin{thm}
For fixed $\alpha$ and $\beta$, $\ckop{\alpha}{\beta}$ can be broken
down into a product of contributions from distinct vertices.
Explicitly,
\bge
   \ckop{\alpha}{\beta} =
   \prod_{\bfx \in L_\alpha}\cvop{\alpha_\bfx}{\beta_\bfx},
   \label{eq:dv}
\ee
where the correlation vertex coefficients (CVC's), ${\cal V}$, are
defined by \label{pg:br}
\bge
   \cvop{\alpha_\bfx}{\beta_\bfx} =
   \sum_{\mu, \nu}
   (-1)^{|\alpha_\bfx \setminus \mu|}
   \left(
      \prod_{i \in (\alpha_\bfx \setminus \mu)} \nnsupo{i}{0}
   \right)
   \left(
      \prod_{j \in (\nu \setminus \beta_\bfx)} \nnsupo{j}{0}
   \right) \vop{\mu}{\nu},
   \label{eq:vdef}
\ee
with the sum taken over $\mu \subseteq \alpha_\bfx$ and $\nu \supseteq
\beta_\bfx,
\nu \subseteq B$.
\end{thm}
This theorem can be proven by expressing $\left.\pdv f^\alpha / \pdv
\Gamma^\beta \right|_0$ and $\left.\pdv g^\alpha / \pdv
\nnsup{\beta}\right|_0$ as products of the $\nnsupo{i}{0}$ and plugging
into (\ref{eq:dexp}).  For a full derivation see \cite{bw1}.  Using this
result, it is possible to express every term in $\cjop{a}{b}$ in
diagrammatic form.  A general nonzero term is given by
\bge
\ckop{\alpsupp{k}}{\betsupp{k}}
\advop{\betsupp{k}}{\alpsupp{k-1}}
\ckop{\alpsupp{k-1}}{\betsupp{k-1}}
\cdots
\advop{\betsupp{2}}{\alpsupp{1}}
\ckop{\alpsupp{1}}{\betsupp{1}}
\label{eq:term}
\ee
with $\alpsupp{i}$ and $\betsupp{i}$ fixed (i.e., not summed over), and
$|\alpsupp{i}|, |\betsupp{i}| \geq 2$, except for the endpoints where
$\alpsupp{k} = \{a\}$ and $\betsupp{1} = \{b\}$.

We define a {\it diagram} $T$ by an integer $k(T)$, which we refer to as
the {\it length} of the diagram $T$, \label{pg:bd} and a function
$\alpha_{T}(\tau)$, where for each $\tau\in\{0,\ldots,k\}$,
$\alpha_{T}(\tau) \subseteq {\cal B}$.  Geometrically, we associate each
$a\in\alpha_{T}(\tau)$ with a {\sl virtual particle} (VP) moving from
$(\bfx (a),\tau)$ to $(\bfx (a) +
\bfcsup{i(a)}, \tau+1)$ on the lattice $\Lambda_{k+1}=
L\times\{0,\ldots,k+1\}$.  We refer to $\alpha_{T}(\tau)$ as the set of
{\sl outgoing} VP's for the diagram $T$. \label{pg:ak}

It is natural to define a corresponding set of {\sl incoming} VP's for
$\tau > 0$ by $\beta_{T}(\tau) = \{b: a(i(b),\bfx (b)-\bfcsup{i(b)}) \in
\alpha_{T}(\tau-1)\}.$ We also define $\sigma_{T}(\tau) = |\alpha_{T}(\tau)|$
to be the total number of outgoing VP's for each value of $\tau$.
Finally, given a diagram $T$, we can define a weight function
\[
   W(T) = \prod_{\bfx \in L} \prod_{1 \leq \tau \leq k(T)}
          \cvop{\alpha_{T}(\tau)_\bfx}{\beta_{T}( \tau)_\bfx},
\]
by taking the product of ${\cal V}$ over all vertices.

The term (\ref{eq:term}) can now be represented by the diagram $T$ with
$\alpha_{T}(\tau) =\alpsupp{\tau}$, where for consistency $\alpsupp{0}$ is
defined to be the unique set with $\advop{\betsupp{1}}{\alpsupp{0}}=1$.
When $\tau \neq 0$, $\advop{\betsupp{\tau+1}}{\alpsupp{\tau}} = 1$, so
$\beta_T(\tau) = \betsupp{\tau}$ for all $\tau$.
It follows that the contribution from (\ref{eq:term}) is exactly given
by  $W(T)$.
Thus, we can rewrite expression (\ref{eq:rj}) for
$\cjop{a}{b}$ as a sum over diagrams
\bge
   \cjop{a}{b} = \sum_{k=1}^\infty \sum_{T\in \ctop{a}{b}(k)} W(T),
   \label{eq:caljsum}
\ee
where in general we define the set of diagrams
$\ctop{\alpha}{\beta}(k)$ by
\begin{eqnarray*}
\ctop{\alpha}{\beta}(k)  & = & \{T:k = k (T), \;
\sigma_T (l) > 1, \;\mbox{for}\;1\leq
l < k,  \;
\alpha_{T} (k) = \alpha,\beta_{T} (1) =  \beta\}.
\end{eqnarray*}

Note that $\cvop{\emptyset}{\nu}=\mixten{\delta}{\emptyset}{\nu}$, so
any diagram with incoming VP's at $\bfxt$ but no outgoing VP's has
weight zero.  For many lattice gases certain other vertex factors
$\cvop{\mu}{\nu}$ vanish also; diagrams with such vertices can be
dropped from the sum~(\ref{eq:caljsum}).  From (\ref{eq:rjt}),
$\tilde{J}$ can now be written as a sum over diagrams in the same
fashion as ${\cal J}$,
\bge
   \tjij (\bfx) =  \sum_{k = 1}^{ \infty}
\sum_{T \in \ctop{i}{j}(\bfx,k)} W(T),
   \label{eq:jsum}
\ee
where the set of diagrams to be summed over is given by
\[
\ctop{i}{j}(\bfx, k) =
\bigcup_{b:i (b) = j} \ctop{a (i,\bfx)}{b}(k).
\]

\section{Approximations}
\setcounter{equation}{0}
\label{ssec:approxx}

We have so far managed to write the exact formula for the hydrodynamic
equations in the scaling limit only in terms of an infinite formal
series.  The natural next question to confront is whether this series
can be summed.  We would like to know whether the series is finite,
and if we cannot sum the full series, at least we would like to find a
set of reasonable approximations which we can make to truncate the
series to one which is summable.  The questions of convergence are
rather difficult, and we will not address them here; in general, the
convergence properties of the series depend on the form of the
conserved quantities in the system.  A variety of methods for
performing partial sums of infinite series of diagrams while retaining
physically important terms have been applied to related problems in
quantum field theory and quantum many-body theory~\cite{IZ,FW}.  We
will describe here several particular approximation methods which are
useful for the kind of series which arise for known lattice gases.

\subsection{Short-$\tau$ and Small-$\ell$ Truncations}
The simplest useful approximations involve truncating the sum
(\ref{eq:caljsum}) to a finite number of terms by putting an upper bound
on either the number of timesteps or the number of distinct nontrivial
vertices allowed in each diagram.  In the first case, the expression for
the renormalized $J$-matrix is
\[
   \tjop{(\tau)i}{j} (\bfx)
   = \sum_{k=1}^\tau\sum_{T \in \ctop{i}{j}(\bfx,k)} W(T),
\]
where the diagrams summed over are the same as those summed in
Eq.~(\ref{eq:jsum}).  Since for each fixed value of $k$ there are a
finite number of allowed diagrams, this sum is finite.  We refer to this
approximation as the {\sl short} $\tau$ {\sl approximation}.  In the
second case, we allow $k$ to be arbitrary, but allow only diagrams where
the total number of vertices $(\bfx, k')$ with nonempty outgoing sets
$\alpha_T( k')_\bfx$ is less than or equal to some fixed number $\ell$.
We denote the sum restricted to these diagrams by $\tjop{[\ell]i}{j}$.
Again, there are only a finite number of such diagrams in this sum,
which means that this sum must also be finite.  This approximation is
analogous to the weak-coupling expansions in quantum field theory,
although in this case the coupling constants $\cvop{\mu}{\nu}$ are
usually not particularly small.  The short-$\tau$ and small-$\ell$
truncations give good consecutive approximations for many lattice gases,
and in most instances pick out the contributions to the renormalization
of the transport coefficients in decreasing order.  In either of these
two approximations, the Boltzmann approximation can be recovered, by
taking $\tau = 1$ or $\ell = 1$.

\subsection{BBGKY Truncations}
Another good class of approximations, in which a reduced but still
infinite set of diagrams is summed, corresponds to truncations of the
BBGKY hierarchy of equations.  Such an approximation involves neglecting
$q$-CCF's with $q > n$ for some fixed value of $n$.  In our diagrammatic
formalism, this amounts to restricting the sum to diagrams with
$\sigma_T(k') \leq n$ for $1 \leq k' \leq k$.  Whereas the total number
of distinct diagrams in the complete sum grows exponentially in $k$, the
number of distinct diagrams in the BBGKY approximations grow
polynomially, and are computationally more tractable.

\subsection{The Ring Approximation}
The $n=2$ version of the BBGKY approximation is closely related to the
{\sl ring approximation}.  The ring approximation is made by neglecting
interactions between two propagating correlated quantities except at the
initial and final vertices of a diagram.  It is generally possible to
calculate a closed-form expression for the infinite sum of diagrams
corresponding to this approximation.  Furthermore, it is usually fairly
easy to calculate the asymptotic form of this approximation as $k
\rightarrow\infty$.  This calculation often captures the most
significant part of the long-term renormalization effects.  In
particular, for certain lattice gases which model two dimensional fluid
systems, the ring approximation diverges logarithmically in $|L|$, which
is in agreement with predictions from other theoretical
frameworks~\cite{Ern}, and also with observed behaviour~\cite{Kad}.

\section{The 1D3P Lattice Gas}

As a simple example, we consider a diffusive lattice gas model in one
dimension.  The model has three bits per site ($n=3$), corresponding to
the presence or absence of left-moving, stationary, and right-moving
particles, respectively.  These bits are denoted by the respective
elements of the set $B=\{-,0,+\}$.  Collisions occur only if exactly two
particles enter a site.  If we denote the two-particle states by
$\widehat{+}\equiv\{-,0\}$, $\widehat{0}\equiv\{-,+\}$, and
$\widehat{-}\equiv\{0,+\}$, then the nontrivial elements of the state
transition table can be written
\bgc
\begin{tabular}{|c|c||c|c|c|}
\hline
\multicolumn{2}{|c||}{$a(s\rightarrow s')$} & \multicolumn{3}{c|}{$s'$} \\
\cline{3-5}
\multicolumn{2}{|c||}{} & $\widehat{+}$ &
\raisebox{-1.8pt}{$\widehat{0}$} & \raisebox{-1.8pt}{$\widehat{-}$}\\
\hline
\hline
    & $\widehat{+}$ & $1-\nsup{p}$ & $\nsup{p}(1-\nsup{r})$ &
$\nsup{p}\nsup{r}$ \\
\cline{2-5}
$s$ & \raisebox{-1.8pt}{$\widehat{0}$} & $\nsup{p}\nsup{r}$ &
$1-\nsup{p}$ & $\nsup{p}(1-\nsup{r})$ \\
\cline{2-5}
    & \raisebox{-1.8pt}{$\widehat{-}$}& $\nsup{p}(1-\nsup{r})$ &
$\nsup{p}\nsup{r}$ & $1-\nsup{p}$ \\
\hline
\end{tabular}
\ec
The bits $n^p$ and $n^r$ are stochastic bits\footnote{Note that $r$
and $p$ are not indices here, but rather simply labels for the
stochastic bits.}
\label{pg:bg} which are sampled separately at each lattice site and at
each timestep with average values $\langle\nsup{p}\rangle = 2p$ and
$\langle\nsup{r}\rangle = 1/2$.  Here, the parameter $p\in
[0,\frac{1}{2}]$ may be thought of as the probability of collision from,
e.g., $\widehat{+}$ to $\widehat{0}$.  The value of the bit $\nsup{p}$
effectively determines whether or not a collision will occur, and that
of $\nsup{r}$ determines which of the two possible outcomes will result.

Note that these collisions obey semi-detailed balance, since the columns
of the above table sum to unity.  Also note that they conserve particles
($n_c=1$); it is the particle density $f$ that will obey the macroscopic
diffusion equation.

Beginning with the above state transition table, we can calculate the
mean vertex coefficients $\mixten{V}{\mu}{\nu}$ using equation
(\ref{eq:vertexmean1}) or (\ref{eq:vertexmean2}).  The nonzero mean
vertex coefficients are given by
\begin{eqnarray*}
\mixten{V}{B}{B} & = & \mixten{V}{\emptyset}{\emptyset} = 1\\
\mixten{V}{i}{j} & = & \mixten{\delta}{i}{j}\\
\mixten{V}{i}{\widehat{j}} & = &  p (3\mixten{\delta}{i}{j}- 1)\\
\mixten{V}{\widehat{i}}{ \widehat{j}} & = &  p +\mixten{\delta}{i}{j} (1 - 3p).
\end{eqnarray*}
Note that since the ensemble-averaged collision operator is invariant
under permutations (relabeling) on the bits, the mean vertex
coefficients also have this symmetry.

\begin{figure}
\centering
\begin{picture}(350,350)(0,0)
\put(70.,306.25){
\begin{picture}(70,70)(0,0)
\put(35.,52.5){\vector(0,-1){15.}}
\put(35.,35.){\vector(0,-1){17.5}}
\put(35.,35.){\circle*{5.}}
\put(35.,0.){\makebox(0,0){\footnotesize $\cvop{0}{0} = 1-2pf$}}
\end{picture}}
\put(210.,306.25){
\begin{picture}(70,70)(0,0)
\put(35.,52.5){\vector(0,-1){15.}}
\put(35.,35.){\vector(-1,-1){17.5}}
\put(35.,35.){\circle*{5.}}
\put(35.,0.){\makebox(0,0){\footnotesize $\cvop{-}{0} = pf$}}
\end{picture}}
\put(0.,218.75){
\begin{picture}(70,70)(0,0)
\put(35.,52.5){\vector(0,-1){15.}}
\put(35.,35.){\vector(-1,-1){17.5}}
\put(35.,35.){\vector(0,-1){17.5}}
\put(35.,35.){\circle*{5.}}
\put(35.,0.){\makebox(0,0){\footnotesize $\cvop{\widehat{+}}{0} = -pf(1-f)$}}
\end{picture}}
\put(140.,218.75){
\begin{picture}(70,70)(0,0)
\put(35.,52.5){\vector(0,-1){15.}}
\put(35.,35.){\vector(-1,-1){17.5}}
\put(35.,35.){\vector(1,-1){17.5}}
\put(35.,35.){\circle*{5.}}
\put(35.,0.){\makebox(0,0){\footnotesize $\cvop{\widehat{0}}{0} = 2pf(1-f)$}}
\end{picture}}
\put(280.,218.75){
\begin{picture}(70,70)(0,0)
\put(17.5,52.5){\vector(1,-1){15.7322}}
\put(35.,52.5){\vector(0,-1){15.}}
\put(35.,35.){\vector(0,-1){17.5}}
\put(35.,35.){\circle*{5.}}
\put(35.,0.){\makebox(0,0){\footnotesize $\cvop{0}{\widehat{-}} = -p$}}
\end{picture}}
\put(0.,131.25){
\begin{picture}(70,70)(0,0)
\put(17.5,52.5){\vector(1,-1){15.7322}}
\put(35.,52.5){\vector(0,-1){15.}}
\put(35.,35.){\vector(-1,-1){17.5}}
\put(35.,35.){\circle*{5.}}
\put(35.,0.){\makebox(0,0){\footnotesize $\cvop{-}{\widehat{-}} = 2p$}}
\end{picture}}
\put(140.,131.25){
\begin{picture}(70,70)(0,0)
\put(17.5,52.5){\vector(1,-1){15.7322}}
\put(35.,52.5){\vector(0,-1){15.}}
\put(35.,35.){\vector(-1,-1){17.5}}
\put(35.,35.){\vector(0,-1){17.5}}
\put(35.,35.){\circle*{5.}}
\put(35.,0.){\makebox(0,0)
{\footnotesize $\cvop{\widehat{+}}{\widehat{-}} = p(1-f)$}}
\end{picture}}
\put(280.,131.25){
\begin{picture}(70,70)(0,0)
\put(17.5,52.5){\vector(1,-1){15.7322}}
\put(35.,52.5){\vector(0,-1){15.}}
\put(35.,35.){\vector(0,-1){17.5}}
\put(35.,35.){\vector(1,-1){17.5}}
\put(35.,35.){\circle*{5.}}
\put(35.,0.){\makebox(0,0)
{\footnotesize $\cvop{\widehat{-}}{\widehat{-}} = 1-2p(1-f)$}}
\end{picture}}
\put(140.,43.75){
\begin{picture}(70,70)(0,0)
\put(17.5,52.5){\vector(1,-1){15.7322}}
\put(35.,52.5){\vector(0,-1){15.}}
\put(52.5,52.5){\vector(-1,-1){15.7322}}
\put(35.,35.){\vector(-1,-1){17.5}}
\put(35.,35.){\vector(0,-1){17.5}}
\put(35.,35.){\vector(1,-1){17.5}}
\put(35.,35.){\circle*{5.}}
\put(35.,0.){\makebox(0,0){\footnotesize $\cvop{B}{B} = 1$}}
\end{picture}}
\end{picture}
\caption{Vertex Factors for 1D3P Lattice Gas}
\label{fig:1d3pvertices}
\end{figure}
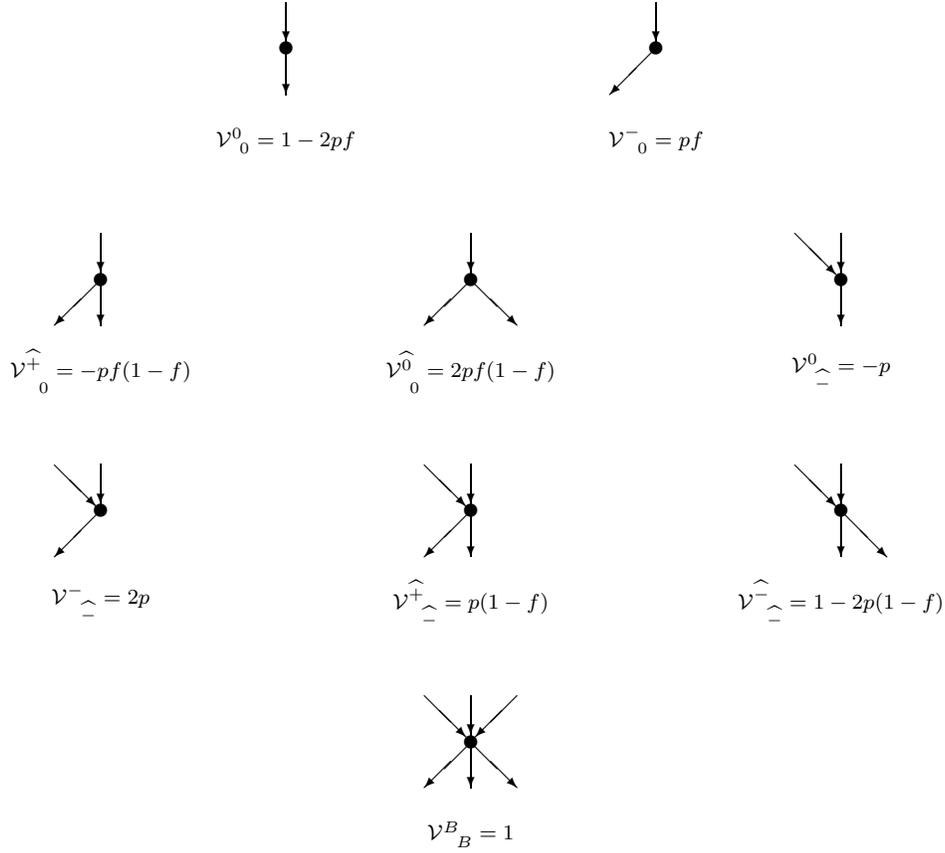

The nonvanishing correlation vertex factors are depicted graphically in
Figure~\ref{fig:1d3pvertices}; only a single vertex is shown in each
equivalence class under the permutation symmetry.  Note that the CVC's
are also symmetric under an arbitrary permutation on the particle
labels.

The Chapman-Enskog analysis for this lattice gas is straightforward, and
yields the following expression for the diffusivity in terms of the
kinetic eigenvalue, $\supscrpt{\tilde{\lambda}}{2}$:
\bge
  \tilde{D} = \frac{c^2}{3\dt}
  \left(\frac{2}{-\supscrpt{\tilde{\lambda}}{2}} -1 \right).
  \label{eq:edoodtplga}
\ee
The Boltzmann value for $\supscrpt{\tilde{\lambda}}{2}$ is $-3pf$.  The
renormalized result can be written
\bge
   \supscrpt{\tilde{\lambda}}{2} = - 3p f + 9 p^2 f (1- f)
   \left(\sum_{\cdots}
\begin{picture}(50,50)(0,22)
\put(35.,50.){\vector(0,-1){17.5}}
\put(35.,50.){\vector(-1,-1){18.2322}}
\put(25.,25.){\makebox(0,0){$\cdots$}}
\put(15.,20.){\vector(0,-1){17.5}}
\put(35.,20.){\vector(-1,-1){18.2322}}
\put(35.,50.){\circle*{5.}}
\put(35.,30.){\circle*{5.}}
\put(15.,30.){\circle*{5.}}
\put(15.,20.){\circle*{5.}}
\put(35.,20.){\circle*{5.}}
\put(15.,0.){\circle*{5.}}
\end{picture}
-\sum_{\cdots}
\begin{picture}(50,50)(0,22)
\put(35.,50.){\vector(0,-1){17.5}}
\put(35.,50.){\vector(-1,-1){18.2322}}
\put(25.,25.){\makebox(0,0){$\cdots$}}
\put(35.,20.){\vector(0,-1){17.5}}
\put(15.,20.){\vector(1,-1){18.2322}}
\put(35.,50.){\circle*{5.}}
\put(35.,30.){\circle*{5.}}
\put(15.,30.){\circle*{5.}}
\put(15.,20.){\circle*{5.}}
\put(35.,20.){\circle*{5.}}
\put(35.,0.){\circle*{5.}}
\end{picture}
   \right),
   \label{eq:1D3Pcorrection}
\ee
where the notation in brackets indicates summation of the products of
all internal vertex factors over all diagrams with the depicted initial
and final configurations.  Together, Eqs.~(\ref{eq:edoodtplga}) and
(\ref{eq:1D3Pcorrection}), with vertices given in
Figure~\ref{fig:1d3pvertices}, constitute an {\sl exact} expression for
the diffusivity of the 1D3P lattice gas.

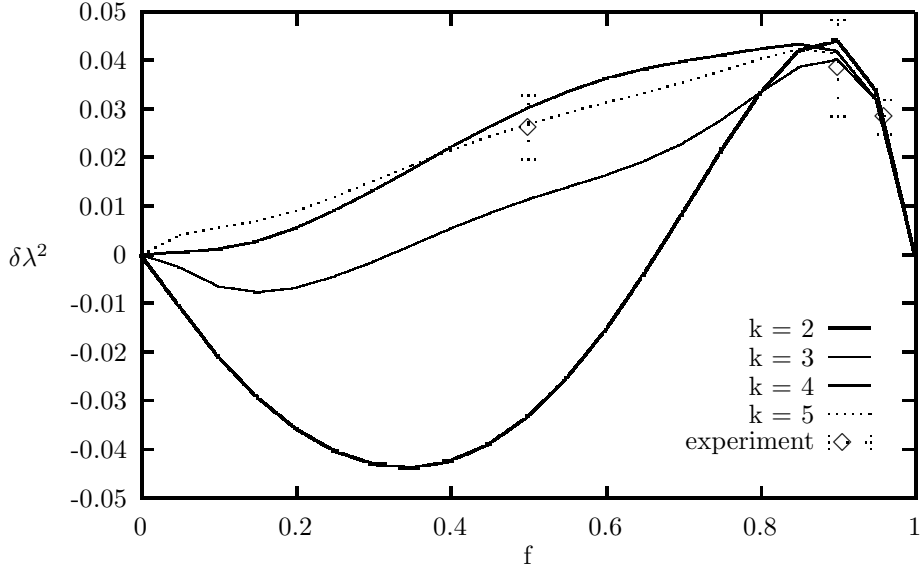
\begin{figure}
\centering
\setlength{\unitlength}{0.240900pt}
\ifx\plotpoint\undefined\newsavebox{\plotpoint}\fi
\sbox{\plotpoint}{\rule[-0.500pt]{1.000pt}{1.000pt}}%
\begin{picture}(1500,900)(0,0)
\font\gnuplot=cmr10 at 10pt
\gnuplot
\sbox{\plotpoint}{\rule[-0.500pt]{1.000pt}{1.000pt}}%
\put(220.0,113.0){\rule[-0.500pt]{4.818pt}{1.000pt}}
\put(198,113){\makebox(0,0)[r]{-0.05}}
\put(1416.0,113.0){\rule[-0.500pt]{4.818pt}{1.000pt}}
\put(220.0,189.0){\rule[-0.500pt]{4.818pt}{1.000pt}}
\put(198,189){\makebox(0,0)[r]{-0.04}}
\put(1416.0,189.0){\rule[-0.500pt]{4.818pt}{1.000pt}}
\put(220.0,266.0){\rule[-0.500pt]{4.818pt}{1.000pt}}
\put(198,266){\makebox(0,0)[r]{-0.03}}
\put(1416.0,266.0){\rule[-0.500pt]{4.818pt}{1.000pt}}
\put(220.0,342.0){\rule[-0.500pt]{4.818pt}{1.000pt}}
\put(198,342){\makebox(0,0)[r]{-0.02}}
\put(1416.0,342.0){\rule[-0.500pt]{4.818pt}{1.000pt}}
\put(220.0,419.0){\rule[-0.500pt]{4.818pt}{1.000pt}}
\put(198,419){\makebox(0,0)[r]{-0.01}}
\put(1416.0,419.0){\rule[-0.500pt]{4.818pt}{1.000pt}}
\put(220.0,495.0){\rule[-0.500pt]{4.818pt}{1.000pt}}
\put(198,495){\makebox(0,0)[r]{0}}
\put(1416.0,495.0){\rule[-0.500pt]{4.818pt}{1.000pt}}
\put(220.0,571.0){\rule[-0.500pt]{4.818pt}{1.000pt}}
\put(198,571){\makebox(0,0)[r]{0.01}}
\put(1416.0,571.0){\rule[-0.500pt]{4.818pt}{1.000pt}}
\put(220.0,648.0){\rule[-0.500pt]{4.818pt}{1.000pt}}
\put(198,648){\makebox(0,0)[r]{0.02}}
\put(1416.0,648.0){\rule[-0.500pt]{4.818pt}{1.000pt}}
\put(220.0,724.0){\rule[-0.500pt]{4.818pt}{1.000pt}}
\put(198,724){\makebox(0,0)[r]{0.03}}
\put(1416.0,724.0){\rule[-0.500pt]{4.818pt}{1.000pt}}
\put(220.0,801.0){\rule[-0.500pt]{4.818pt}{1.000pt}}
\put(198,801){\makebox(0,0)[r]{0.04}}
\put(1416.0,801.0){\rule[-0.500pt]{4.818pt}{1.000pt}}
\put(220.0,877.0){\rule[-0.500pt]{4.818pt}{1.000pt}}
\put(198,877){\makebox(0,0)[r]{0.05}}
\put(1416.0,877.0){\rule[-0.500pt]{4.818pt}{1.000pt}}
\put(220.0,113.0){\rule[-0.500pt]{1.000pt}{4.818pt}}
\put(220,68){\makebox(0,0){0}}
\put(220.0,857.0){\rule[-0.500pt]{1.000pt}{4.818pt}}
\put(463.0,113.0){\rule[-0.500pt]{1.000pt}{4.818pt}}
\put(463,68){\makebox(0,0){0.2}}
\put(463.0,857.0){\rule[-0.500pt]{1.000pt}{4.818pt}}
\put(706.0,113.0){\rule[-0.500pt]{1.000pt}{4.818pt}}
\put(706,68){\makebox(0,0){0.4}}
\put(706.0,857.0){\rule[-0.500pt]{1.000pt}{4.818pt}}
\put(950.0,113.0){\rule[-0.500pt]{1.000pt}{4.818pt}}
\put(950,68){\makebox(0,0){0.6}}
\put(950.0,857.0){\rule[-0.500pt]{1.000pt}{4.818pt}}
\put(1193.0,113.0){\rule[-0.500pt]{1.000pt}{4.818pt}}
\put(1193,68){\makebox(0,0){0.8}}
\put(1193.0,857.0){\rule[-0.500pt]{1.000pt}{4.818pt}}
\put(1436.0,113.0){\rule[-0.500pt]{1.000pt}{4.818pt}}
\put(1436,68){\makebox(0,0){1}}
\put(1436.0,857.0){\rule[-0.500pt]{1.000pt}{4.818pt}}
\put(220.0,113.0){\rule[-0.500pt]{292.934pt}{1.000pt}}
\put(1436.0,113.0){\rule[-0.500pt]{1.000pt}{184.048pt}}
\put(220.0,877.0){\rule[-0.500pt]{292.934pt}{1.000pt}}
\put(45,495){\makebox(0,0){$\delta\lambda^2$}}
\put(828,23){\makebox(0,0){f}}
\put(220.0,113.0){\rule[-0.500pt]{1.000pt}{184.048pt}}
\put(1278,380){\makebox(0,0)[r]{k = 2}}
\put(1300.0,380.0){\rule[-0.500pt]{15.899pt}{1.000pt}}
\put(220,495){\usebox{\plotpoint}}
\multiput(221.83,488.45)(0.499,-0.659){114}{\rule{0.120pt}{1.578pt}}
\multiput(217.92,491.73)(61.000,-77.725){2}{\rule{1.000pt}{0.789pt}}
\multiput(282.83,407.59)(0.499,-0.642){114}{\rule{0.120pt}{1.545pt}}
\multiput(278.92,410.79)(61.000,-75.793){2}{\rule{1.000pt}{0.773pt}}
\multiput(343.83,329.60)(0.499,-0.519){112}{\rule{0.120pt}{1.300pt}}
\multiput(339.92,332.30)(60.000,-60.302){2}{\rule{1.000pt}{0.650pt}}
\multiput(402.00,269.68)(0.603,-0.498){92}{\rule{1.470pt}{0.120pt}}
\multiput(402.00,269.92)(57.949,-50.000){2}{\rule{0.735pt}{1.000pt}}
\multiput(463.00,219.68)(0.864,-0.498){62}{\rule{1.993pt}{0.120pt}}
\multiput(463.00,219.92)(56.864,-35.000){2}{\rule{0.996pt}{1.000pt}}
\multiput(524.00,184.68)(1.452,-0.496){34}{\rule{3.155pt}{0.119pt}}
\multiput(524.00,184.92)(54.452,-21.000){2}{\rule{1.577pt}{1.000pt}}
\multiput(585.00,163.69)(5.920,-0.462){4}{\rule{10.417pt}{0.111pt}}
\multiput(585.00,163.92)(39.380,-6.000){2}{\rule{5.208pt}{1.000pt}}
\multiput(646.00,161.83)(2.794,0.489){14}{\rule{5.705pt}{0.118pt}}
\multiput(646.00,157.92)(48.160,11.000){2}{\rule{2.852pt}{1.000pt}}
\multiput(706.00,172.83)(1.124,0.497){46}{\rule{2.509pt}{0.120pt}}
\multiput(706.00,168.92)(55.792,27.000){2}{\rule{1.255pt}{1.000pt}}
\multiput(767.00,199.83)(0.702,0.498){78}{\rule{1.669pt}{0.120pt}}
\multiput(767.00,195.92)(57.537,43.000){2}{\rule{0.834pt}{1.000pt}}
\multiput(828.00,242.83)(0.502,0.499){112}{\rule{1.267pt}{0.120pt}}
\multiput(828.00,238.92)(58.371,60.000){2}{\rule{0.633pt}{1.000pt}}
\multiput(890.83,301.00)(0.499,0.609){114}{\rule{0.120pt}{1.480pt}}
\multiput(886.92,301.00)(61.000,71.929){2}{\rule{1.000pt}{0.740pt}}
\multiput(951.83,376.00)(0.499,0.729){112}{\rule{0.120pt}{1.717pt}}
\multiput(947.92,376.00)(60.000,84.437){2}{\rule{1.000pt}{0.858pt}}
\multiput(1011.83,464.00)(0.499,0.783){114}{\rule{0.120pt}{1.824pt}}
\multiput(1007.92,464.00)(61.000,92.215){2}{\rule{1.000pt}{0.912pt}}
\multiput(1072.83,560.00)(0.499,0.807){114}{\rule{0.120pt}{1.873pt}}
\multiput(1068.92,560.00)(61.000,95.113){2}{\rule{1.000pt}{0.936pt}}
\multiput(1133.83,659.00)(0.499,0.733){114}{\rule{0.120pt}{1.725pt}}
\multiput(1129.92,659.00)(61.000,86.419){2}{\rule{1.000pt}{0.863pt}}
\multiput(1194.83,749.00)(0.499,0.543){114}{\rule{0.120pt}{1.348pt}}
\multiput(1190.92,749.00)(61.000,64.201){2}{\rule{1.000pt}{0.674pt}}
\multiput(1254.00,817.83)(1.887,0.494){24}{\rule{4.000pt}{0.119pt}}
\multiput(1254.00,813.92)(51.698,16.000){2}{\rule{2.000pt}{1.000pt}}
\multiput(1315.83,825.59)(0.499,-0.642){114}{\rule{0.120pt}{1.545pt}}
\multiput(1311.92,828.79)(61.000,-75.793){2}{\rule{1.000pt}{0.773pt}}
\multiput(1376.83,734.41)(0.499,-2.119){114}{\rule{0.120pt}{4.480pt}}
\multiput(1372.92,743.70)(61.000,-248.703){2}{\rule{1.000pt}{2.240pt}}
\sbox{\plotpoint}{\rule[-0.175pt]{0.350pt}{0.350pt}}%
\put(1278,335){\makebox(0,0)[r]{k = 3}}
\put(1300.0,335.0){\rule[-0.175pt]{15.899pt}{0.350pt}}
\put(220,495){\usebox{\plotpoint}}
\multiput(220.00,494.02)(1.561,-0.501){37}{\rule{1.155pt}{0.121pt}}
\multiput(220.00,494.27)(58.603,-20.000){2}{\rule{0.578pt}{0.350pt}}
\multiput(281.00,474.02)(1.030,-0.501){57}{\rule{0.799pt}{0.121pt}}
\multiput(281.00,474.27)(59.341,-30.000){2}{\rule{0.400pt}{0.350pt}}
\multiput(342.00,444.02)(3.568,-0.503){15}{\rule{2.421pt}{0.121pt}}
\multiput(342.00,444.27)(54.975,-9.000){2}{\rule{1.210pt}{0.350pt}}
\multiput(402.00,436.47)(4.798,0.504){11}{\rule{3.138pt}{0.121pt}}
\multiput(402.00,435.27)(54.488,7.000){2}{\rule{1.569pt}{0.350pt}}
\multiput(463.00,443.48)(1.741,0.501){33}{\rule{1.274pt}{0.121pt}}
\multiput(463.00,442.27)(58.357,18.000){2}{\rule{0.637pt}{0.350pt}}
\multiput(524.00,461.48)(1.352,0.501){43}{\rule{1.016pt}{0.121pt}}
\multiput(524.00,460.27)(58.892,23.000){2}{\rule{0.508pt}{0.350pt}}
\multiput(585.00,484.48)(1.192,0.501){49}{\rule{0.909pt}{0.121pt}}
\multiput(585.00,483.27)(59.114,26.000){2}{\rule{0.454pt}{0.350pt}}
\multiput(646.00,510.48)(1.173,0.501){49}{\rule{0.895pt}{0.121pt}}
\multiput(646.00,509.27)(58.142,26.000){2}{\rule{0.448pt}{0.350pt}}
\multiput(706.00,536.48)(1.294,0.501){45}{\rule{0.977pt}{0.121pt}}
\multiput(706.00,535.27)(58.972,24.000){2}{\rule{0.489pt}{0.350pt}}
\multiput(767.00,560.48)(1.415,0.501){41}{\rule{1.058pt}{0.121pt}}
\multiput(767.00,559.27)(58.804,22.000){2}{\rule{0.529pt}{0.350pt}}
\multiput(828.00,582.48)(1.646,0.501){35}{\rule{1.211pt}{0.121pt}}
\multiput(828.00,581.27)(58.486,19.000){2}{\rule{0.606pt}{0.350pt}}
\multiput(889.00,601.48)(1.646,0.501){35}{\rule{1.211pt}{0.121pt}}
\multiput(889.00,600.27)(58.486,19.000){2}{\rule{0.606pt}{0.350pt}}
\multiput(950.00,620.48)(1.392,0.501){41}{\rule{1.042pt}{0.121pt}}
\multiput(950.00,619.27)(57.837,22.000){2}{\rule{0.521pt}{0.350pt}}
\multiput(1010.00,642.48)(1.105,0.501){53}{\rule{0.850pt}{0.121pt}}
\multiput(1010.00,641.27)(59.236,28.000){2}{\rule{0.425pt}{0.350pt}}
\multiput(1071.00,670.48)(0.832,0.501){71}{\rule{0.665pt}{0.121pt}}
\multiput(1071.00,669.27)(59.621,37.000){2}{\rule{0.332pt}{0.350pt}}
\multiput(1132.00,707.48)(0.714,0.501){83}{\rule{0.584pt}{0.121pt}}
\multiput(1132.00,706.27)(59.788,43.000){2}{\rule{0.292pt}{0.350pt}}
\multiput(1193.00,750.48)(0.769,0.501){77}{\rule{0.621pt}{0.121pt}}
\multiput(1193.00,749.27)(59.711,40.000){2}{\rule{0.311pt}{0.350pt}}
\multiput(1254.00,790.48)(2.619,0.502){21}{\rule{1.837pt}{0.121pt}}
\multiput(1254.00,789.27)(56.186,12.000){2}{\rule{0.919pt}{0.350pt}}
\multiput(1314.48,800.11)(0.500,-0.526){119}{\rule{0.121pt}{0.455pt}}
\multiput(1313.27,801.06)(61.000,-63.056){2}{\rule{0.350pt}{0.227pt}}
\multiput(1375.48,731.85)(0.500,-2.007){119}{\rule{0.121pt}{1.482pt}}
\multiput(1374.27,734.92)(61.000,-239.925){2}{\rule{0.350pt}{0.741pt}}
\sbox{\plotpoint}{\rule[-0.300pt]{0.600pt}{0.600pt}}%
\put(1278,290){\makebox(0,0)[r]{k = 4}}
\put(1300.0,290.0){\rule[-0.300pt]{15.899pt}{0.600pt}}
\put(220,495){\usebox{\plotpoint}}
\multiput(220.00,495.99)(11.197,0.503){3}{\rule{9.300pt}{0.121pt}}
\multiput(220.00,493.75)(41.697,4.000){2}{\rule{4.650pt}{0.600pt}}
\multiput(281.00,499.99)(7.498,0.502){5}{\rule{7.470pt}{0.121pt}}
\multiput(281.00,497.75)(45.496,5.000){2}{\rule{3.735pt}{0.600pt}}
\multiput(342.00,505.00)(2.607,0.500){19}{\rule{3.150pt}{0.121pt}}
\multiput(342.00,502.75)(53.462,12.000){2}{\rule{1.575pt}{0.600pt}}
\multiput(402.00,517.00)(1.476,0.500){37}{\rule{1.893pt}{0.121pt}}
\multiput(402.00,514.75)(57.071,21.000){2}{\rule{0.946pt}{0.600pt}}
\multiput(463.00,538.00)(1.099,0.500){51}{\rule{1.457pt}{0.120pt}}
\multiput(463.00,535.75)(57.976,28.000){2}{\rule{0.729pt}{0.600pt}}
\multiput(524.00,566.00)(0.991,0.500){57}{\rule{1.331pt}{0.120pt}}
\multiput(524.00,563.75)(58.238,31.000){2}{\rule{0.665pt}{0.600pt}}
\multiput(585.00,597.00)(0.902,0.500){63}{\rule{1.226pt}{0.120pt}}
\multiput(585.00,594.75)(58.454,34.000){2}{\rule{0.613pt}{0.600pt}}
\multiput(646.00,631.00)(0.887,0.500){63}{\rule{1.209pt}{0.120pt}}
\multiput(646.00,628.75)(57.491,34.000){2}{\rule{0.604pt}{0.600pt}}
\multiput(706.00,665.00)(0.959,0.500){59}{\rule{1.294pt}{0.120pt}}
\multiput(706.00,662.75)(58.315,32.000){2}{\rule{0.647pt}{0.600pt}}
\multiput(767.00,697.00)(1.024,0.500){55}{\rule{1.370pt}{0.120pt}}
\multiput(767.00,694.75)(58.156,30.000){2}{\rule{0.685pt}{0.600pt}}
\multiput(828.00,727.00)(1.234,0.500){45}{\rule{1.614pt}{0.120pt}}
\multiput(828.00,724.75)(57.650,25.000){2}{\rule{0.807pt}{0.600pt}}
\multiput(889.00,752.00)(1.476,0.500){37}{\rule{1.893pt}{0.121pt}}
\multiput(889.00,749.75)(57.071,21.000){2}{\rule{0.946pt}{0.600pt}}
\multiput(950.00,773.00)(2.059,0.500){25}{\rule{2.550pt}{0.121pt}}
\multiput(950.00,770.75)(54.707,15.000){2}{\rule{1.275pt}{0.600pt}}
\multiput(1010.00,788.00)(2.651,0.500){19}{\rule{3.200pt}{0.121pt}}
\multiput(1010.00,785.75)(54.358,12.000){2}{\rule{1.600pt}{0.600pt}}
\multiput(1071.00,799.99)(3.227,0.501){15}{\rule{3.810pt}{0.121pt}}
\multiput(1071.00,797.75)(53.092,10.000){2}{\rule{1.905pt}{0.600pt}}
\multiput(1132.00,809.99)(3.227,0.501){15}{\rule{3.810pt}{0.121pt}}
\multiput(1132.00,807.75)(53.092,10.000){2}{\rule{1.905pt}{0.600pt}}
\multiput(1193.00,819.99)(4.834,0.501){9}{\rule{5.379pt}{0.121pt}}
\multiput(1193.00,817.75)(49.837,7.000){2}{\rule{2.689pt}{0.600pt}}
\multiput(1254.00,824.50)(2.607,-0.500){19}{\rule{3.150pt}{0.121pt}}
\multiput(1254.00,824.75)(53.462,-12.000){2}{\rule{1.575pt}{0.600pt}}
\multiput(1315.00,810.32)(0.500,-0.615){117}{\rule{0.120pt}{0.888pt}}
\multiput(1312.75,812.16)(61.000,-73.158){2}{\rule{0.600pt}{0.444pt}}
\multiput(1376.00,728.41)(0.500,-2.012){117}{\rule{0.120pt}{2.550pt}}
\multiput(1373.75,733.71)(61.000,-238.707){2}{\rule{0.600pt}{1.275pt}}
\sbox{\plotpoint}{\rule[-0.250pt]{0.500pt}{0.500pt}}%
\put(1278,245){\makebox(0,0)[r]{k = 5}}
\multiput(1300,245)(12.453,0.000){6}{\usebox{\plotpoint}}
\put(1366,245){\usebox{\plotpoint}}
\put(220,495){\usebox{\plotpoint}}
\multiput(220,495)(11.102,5.642){6}{\usebox{\plotpoint}}
\multiput(281,526)(12.219,2.404){5}{\usebox{\plotpoint}}
\multiput(342,538)(12.284,2.047){5}{\usebox{\plotpoint}}
\multiput(402,548)(12.046,3.160){5}{\usebox{\plotpoint}}
\multiput(463,564)(11.715,4.225){5}{\usebox{\plotpoint}}
\multiput(524,586)(11.523,4.723){5}{\usebox{\plotpoint}}
\multiput(585,611)(11.523,4.723){6}{\usebox{\plotpoint}}
\multiput(646,636)(11.628,4.457){5}{\usebox{\plotpoint}}
\multiput(706,659)(11.715,4.225){5}{\usebox{\plotpoint}}
\multiput(767,681)(11.890,3.703){5}{\usebox{\plotpoint}}
\multiput(828,700)(11.944,3.525){5}{\usebox{\plotpoint}}
\multiput(889,718)(12.046,3.160){5}{\usebox{\plotpoint}}
\multiput(950,734)(12.081,3.020){5}{\usebox{\plotpoint}}
\multiput(1010,749)(11.996,3.343){5}{\usebox{\plotpoint}}
\multiput(1071,766)(11.944,3.525){6}{\usebox{\plotpoint}}
\multiput(1132,784)(11.890,3.703){5}{\usebox{\plotpoint}}
\multiput(1193,803)(12.093,2.974){5}{\usebox{\plotpoint}}
\multiput(1254,818)(12.369,-1.443){5}{\usebox{\plotpoint}}
\multiput(1314,811)(8.050,-9.502){7}{\usebox{\plotpoint}}
\multiput(1375,739)(3.020,-12.081){20}{\usebox{\plotpoint}}
\put(1436,495){\usebox{\plotpoint}}
\put(1278,200){\makebox(0,0)[r]{experiment}}
\put(1322,200){\raisebox{-.8pt}{\makebox(0,0){$\Diamond$}}}
\put(828,695){\raisebox{-.8pt}{\makebox(0,0){$\Diamond$}}}
\put(1314,788){\raisebox{-.8pt}{\makebox(0,0){$\Diamond$}}}
\put(1387,711){\raisebox{-.8pt}{\makebox(0,0){$\Diamond$}}}
\multiput(1300,200)(29.058,0.000){3}{\usebox{\plotpoint}}
\put(1366,200){\usebox{\plotpoint}}
\put(1300.00,210.00){\usebox{\plotpoint}}
\put(1300,190){\usebox{\plotpoint}}
\put(1366.00,210.00){\usebox{\plotpoint}}
\put(1366,190){\usebox{\plotpoint}}
\multiput(828,645)(0.000,29.058){4}{\usebox{\plotpoint}}
\put(828,745){\usebox{\plotpoint}}
\put(818.00,645.00){\usebox{\plotpoint}}
\put(838,645){\usebox{\plotpoint}}
\put(818.00,745.00){\usebox{\plotpoint}}
\put(838,745){\usebox{\plotpoint}}
\multiput(1314,712)(0.000,29.058){6}{\usebox{\plotpoint}}
\put(1314,864){\usebox{\plotpoint}}
\put(1304.00,712.00){\usebox{\plotpoint}}
\put(1324,712){\usebox{\plotpoint}}
\put(1304.00,864.00){\usebox{\plotpoint}}
\put(1324,864){\usebox{\plotpoint}}
\multiput(1387,684)(0.000,29.058){2}{\usebox{\plotpoint}}
\put(1387,738){\usebox{\plotpoint}}
\put(1377.00,684.00){\usebox{\plotpoint}}
\put(1397,684){\usebox{\plotpoint}}
\put(1377.00,738.00){\usebox{\plotpoint}}
\put(1397,738){\usebox{\plotpoint}}
\end{picture}
\caption{Partial BBGKY Approximations for $p=1/2$}
\label{fig:bbgky2}
\end{figure}

We have used a computer to numerically calculate the limit of the full
$k$-particle BBGKY approximation for certain values of $f$ and $p$.
The algorithm we used was to  sum all diagrams of length $\leq \tau$
on a lattice of width $l$, then to take the limits as $\tau, l
\rightarrow \infty$.  In this way, we have numerically approximated the
limits of the $k$-particle BBGKY sums for $k < 6$.  The results of this
calculation, plotted as the {\it correction} to the Boltzmann value of
the kinetic eigenvalue, versus mean occupation number, are graphed in
Figure~\ref{fig:bbgky2} for $p=1/2$.  It is especially noteworthy that
when $f=1/2$ the ring approximation and lower BBGKY truncations are
inadequate (indeed, the $k=2$ and ring truncations actually get the
wrong sign for the correction), but the $k=5$ BBGKY truncation does
quite well indeed.

\section{Conclusions}
\setcounter{equation}{0}

In this paper, we have outlined a complete kinetic theory of lattice
gases, applied it to a simple model lattice gas, and compared the
predictions of the theory to experiment.

The approach presented in this paper opens up a wide range of possible
new work on discrete kinetic theory.  By applying these techniques to
compute deviations from the Boltzmann predictions for commonly used
lattice gases, the results of simulations can be more accurately
interpreted.  Lattice gases are currently being used, both in industrial
and academic settings, for computational fluid dynamics
calculations~\cite{kim}; to ensure the accuracy of these calculations,
it is essential to account for the renormalization effects that we have
studied here.

In addition to quantitative refinement of lattice gas calculations, the
theory presented here provides a tool with which to investigate
fundamental physical phenomena in nonequilibrium statistical systems.
In recent years, for example, lattice gases have been used to model many
different hydrodynamic systems, including reaction-diffusion equations,
and other systems capable of spontaneous self-organization.  It is
known~\cite{lw} that the Boltzmann approximation does not yield accurate
results for the transport coefficients of such systems, unless the
reactants are allowed to diffuse for several steps between reactions in
order to artificially supress the correlations that develop~\cite{dab}.
Thus, it is likely that the methods described in this paper will be
directly applicable to these systems, providing an important correction
to their theory.  More interestingly, these methods will also provide
insight into the extremely subtle flow and agglomeration of
interparticle correlations in pattern-forming lattice gases, and hence
into the dynamical basis of self-organization.

\section*{Acknowledgements}
One of us (BMB) would like to acknowldege helpful conversations with and
encouragement from Professors M.H. Ernst and E.G.D. Cohen, and from Dr.
B. Hasslacher.  In addition, he would like to acknowledge the
hospitality of the Information Mechanics Group at the M.I.T. Laboratory
for Computer Science where he was a visiting scientist during a portion
of this work.  This work was supported in part by Thinking Machines
Corporation, in part by the divisions of Applied Mathematics of the U.S.
Department of Energy (DOE) under contracts DE-FG02-88ER25065 and
DE-FG02-88ER25066, and in part by the U.S. Department of Energy (DOE)
under cooperative agreement DE-FC02-94ER40818.

\end{document}